September 12, 2014

# The Distributed Language

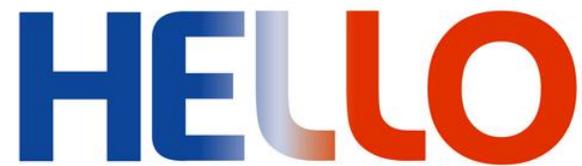

*V1.0.3 (alpha)*

# White Paper

Boris Burshteyn
bburshteyn@amsdec.com

## Abstract

Hello is a general-purpose, object-oriented, protocol-agnostic distributed programming language. This paper explains the ideas that guided design of Hello. It shows the spirit of Hello using two brief expressive programs and provides a summary of language features. In addition, it explores historical parallels between the binary programming of early computers and the distributed programming of modern networks.

Hello documentation, translator and runtime engine are freely available at [14].

## 1  Preface

> *"Whether oratory is a creation of rules, or of training, or of natural gifts, it is the most difficult of all things to achieve."* Cicero, "The Brutus," 46 B.C.E.

Cicero – the most gifted orator of all times – had realized the extreme difficulty of communication more than two thousand years ago. Obviously, there were no computers when he made his famous speeches in the Roman Courts, Assemblies, and Senate, and wrote treatises about the oratory art. At that time, his talents flourished in the area of one-way human communication.

Today computers exchange code and data across the networks with the help of sophisticated hardware devices, intricate networking protocols and elaborate software systems. This immensely complex communication infrastructure makes creation of networking software an extremely difficult and costly enterprise. Thus, the Cicero's quote remains true when applied to modern computer networking – distributed software is one of the hardest fields of software engineering.

The distributed applications orchestrate computations across the network while utilizing a multitude of collaborating computers. Their development demands significant investments in labor, time, and finances. Therefore, a relentless quest for a silver bullet that would reduce the development efforts continues.

This paper presents Hello -- a general-purpose, object-oriented, imperative, protocol-agnostic programming language for writing distributed software. Its unique first-class distributed language features put Hello in an excellent position to cut the costs and ease the pain of the distributed software development.

## 2  Background

The idea to create a distributed programming language emerged during the long experience of designing and implementing commercial-grade networking applications as diverse as a Distributed Database, High Availability Cluster Monitor, MAPI-based Email Server (via wire-level protocol discovery), and a Video-Conferencing Server. During that work, communication libraries implemented abstractions that hid the underlying network protocol. These abstractions represented interconnected servers communicating over the network between themselves and with clients.

## 3  Three Problems

Although the applications were quite different, they abstracted the distributed hardware and software in a similar way. Still, the abstractions had to be implemented each time anew as all systems had re-invented many of the same building blocks: the re-programming involved remote objects, connections, messages, queues, jobs, replies, timeouts, state changes, failures, and others.

Because the language in hand had no first-class distributed features, it needed help of extraneous distributed libraries. However, since the language knew nothing about the libraries, which offered only low level networking interface, the coding and debugging were difficult for programmers while the optimization was impossible for a translator. At the end, there were no time and project resources left to improve efficiency and reliability of the communication components. In summary, the following three problems characterized many distributed systems:

1) *Repeated re-coding of the same distributed primitives.*
2) *Challenging development process due to the combination of a programming language and extraneous library.*
3) *Inefficient and unreliable distributed code.*

## 4  One Solution

Those repetitive efforts had built determination to develop distributed primitives for reuse in a diverse spectrum of applications. To embed them in the language was a logical choice because it offered simultaneous simplicity of thought and efficiency of code[1]. The distributed programming language has nicely solved all three problems mentioned above:

1) Since the first-class distributed language features are always available, there is no need to re-implement them for every new application.
2) These protocol-agnostic features, being higher level than the library interfaces, cut the need for extraneous tools while simplifying development and debugging.
3) The distributed object-oriented language architecture and the optimizing translator raise the efficiency, reliability, and security of the distributed code.

The distributed elements, implemented in both the language translator and its runtime system, become as ubiquitous as are the local primitives like heap, stack, method, class, variable, and object. Therefore, one does not need extraneous libraries, but can rely only on the language translator, and runtime to achieve the productive software development combined with the efficient and reliable distributed code. Thus, the distributed programming language Hello was born to blend the protocol-agnostic network model into the fabric of an object-oriented paradigm.

---

[1] It was also natural because the author had extensive experience with parser generators and a variety of language frontends.

## 5  Language Design Goals

One of the primary goals set during the Hello design was its future widespread adoption by the software engineering community. Therefore, it was crucial to come up with a general-purpose language, avoiding the danger of falling into a niche of specialized languages [20]. Another goal was to abstract in the language the components of the entire network and its protocols. The abstraction level should be high enough to conceal the mundane intricacies of the rigid protocol interfaces, yet not too high to make the language overly specialized. In addition, distributed primitives were meant to become a crucial part of the language, but not its dominant part: the language core should remain filled with general-purpose abstractions for sequential computations, which are running on a single host.

Towards these goals, *the lower-level distributed abstractions embedded into the language's type system, augmented by a few imperative operations, had been invented.* Any particular distributed semantics can now be derived from a combination of the Hello distributed primitives -- engines, hosts, queues, partitions, groups, messages, events, and remote references -- and their operations.

## 6  Language Design Principle

Historically, single-host general-purpose languages abstract the computer memory, memory-stored programs, and CPU from the computer hardware architecture first articulated in 1945 by J. von Neumann in [21]. However, distributed abstractions should differ from single-host abstractions because of the following fundamental differences between the single-host and distributed architectures.

As J. Backus explained in [1], imperative single-host languages are designed so that their programs reflect the linear structure of the von Neumann machine: variables and statements from the source text of a program are translated into data and CPU instructions stored in the contiguous memory of a single host. Moreover, the linear text-descending control flow (with occasional `gotos` and loops) in the source program corresponds to the linear address-ascending control flow (with occasional jumps and branches) in its translated machine code, which is stored in a single piece of memory at runtime.

However, distributed software does not run in a single memory piece – its components execute on different computers with separate memory units. The computers connect, communicate and compute in a network of any structure: from linear – to hierarchical -- to the full graph. Network topology and connections between application's components often change with time. Thus, the linear text-to-memory mapping idea that guided design of the imperative single-host languages fails for distributed languages. At the same time, a distributed language should abstract the networked components such as computers, processes, threads, objects, messages, events, and code packages. It shall provide first-class language features to manage distributed operations with inevitable timeouts and failures, and to hide the underlying networking protocol.

*Therefore, Hello design principle was to implement the distributed abstractions inside the language's type system*:

1) Hello provides several built-in classes augmented with a few operations and statements, organically interwoven into the language's imperative syntax and object-oriented semantics. Instances of these classes represent hosts, runtime engines, memory partitions, and execution queues spread out around the network; their purpose is to support first-class distributed language features transparently to Hello programs. At runtime, host instances self-organize into distributed groups, which graph structures reflect dynamic network connections.
2) The instances of user-defined classes – objects – reside in memory partitions and engine heaps around the network referring to each other through local and remote references. The runtime engines, with the help of the translator and the instances of the built-in classes, transparently navigate object references. The navigation allows Hello programs to manipulate and transfer data, code and control flow over the network.

## 7  Hello, World!

The following example shows how Hello first-class distributed features and sequential core allow for a simple design. In addition, it demonstrates fast coding of the short and efficient distributed programs. This example highlights Hello advantages over the single-host languages C [4], C++ [5] and Java [18] augmented with extraneous distributed libraries.

The program "Hello, World!" shows just some first-class distributed features of the Hello language, which set it apart from any single-host programming language. The whole program is a class named `HelloWorld`. It contains a function `main()` that invokes a function `print()` on the hosts referred to from the built-in group `hosts`:

```
package Hello_World;
class HelloWorld {      // Broadcast to all known hosts
    public static void main() {
        hosts.+print("Hello, World!\n" +
                     this_host.name() + ":-)\n");
    }
};
```

The two Hello concepts illustrated by this example are a *group*, which is a graph-ordered collection of objects, and operator of *bottom-up iteration* `.+`, which invokes, in the graph order, an *iterator function* on each of the group's objects. In particular, the built-in group named `hosts`, has its objects distributed across the network -- each object is a separate instance of class `host_group`, residing within the virtual or physical host, which this object describes. At runtime, Hello engines maintain this group automatically: two objects are connected in the group if their hosts can communicate over the network. The definition of `host_group` is located in the Hello package `standard`; here is its relevant abbreviated fragment:

```
public external group class host_group {
  public external           host         current_host;
  public external copy      host_group[] children();
  public external iterator void print(copy char[] str) {
      current_host.print(str);
  }
};
```

When `main()` is started on a runtime engine, it begins iterating function `print()` on the objects from the group `hosts`. Because each host object resides on a different host, `print()` is invoked on different hosts -- once locally on the current host referred to by `current_host` and once remotely on each of the remote children hosts referred to by references from the array returned by the function with the reserved name `children()`.

During the iteration, at each invocation `print()` writes a greeting signed with the name of the host where `main()` has been started and kept waiting for the iteration to finish; this name is a character array returned by the method `current_host.name()`. At each invocation, `print()` writes to the standard output of the host where `print()` is invoked, local or remote. Its argument is a string combined, with the operator concatenation +, from the three portions of the greeting: two literal strings and one character array. It is copied across the network during iteration and passed to `print()` at each invocation.

Package `standard` is an internal part of the Hello runtime – it contains types that support distributed architecture (e.g. classes `host` and `host_group`). A programmer may want to learn a general functionality of some standard types and operations described in Hello Programming Guide [14]. However, there is no need to understand its details or even import it into any package.

### 7.1 Hello, C++ & Java!

The simple imperative and object-oriented features coupled with the distributed abstractions distinguish general-purpose Hello architecture from the specialized distributed architectures[2]. In addition, the following code snippets present "Hello, World!" broadcasts in Java and C++ as seen at [13] and [3].

Since neither Java nor C++ has distributed features, both resort to extraneous libraries: Java to RMI, C++ to BOOST.mpi. The difference with the "Hello, World!" written in Hello is apparent – the latter has no setup while both C++ and Java use most of their code (in red) for setting up translation and runtime contexts.

At the same time, Hello code is much shorter than either C++ or Java. Moreover, the Java source shows only the server part – the client portion from [13] adds even more Java code while Hello broadcast does not distinguish between client and server.

```
package example.hello;
import java.rmi.registry.Registry;
import java.rmi.registry.LocateRegistry;
import java.rmi.RemoteException;
import java.rmi.server.UnicastRemoteObject;
public class Server implements Hello {
  public Server() {}
  public String sayHello() {
    return "Hello, world!";
  }
  public static void main(String args[]) {
    try {
      Server obj = new Server();
      Hello stub = (Hello)
                UnicastRemoteObject.
                     exportObject(obj, 0);
      // Bind the remote object stub in the registry
      Registry registry = LocateRegistry.
                     getRegistry();
      registry.bind("Hello", stub);
      System.err.println("Server ready");
    } catch (Exception e) {
      System.err.println("Server exception: " +
                e.toString());
      e.printStackTrace();
    }
  }
}
```

**Figure 1 Java broadcasting server program as seen at [13]**

```
#include <boost/mpi.hpp>
#include <iostream>
#include <boost/serialization/string.hpp>
namespace mpi = boost::mpi;
int main(int argc, char* argv[]) {
  mpi::environment env(argc, argv);
  mpi::communicator world;
  std::string value;
  if (world.rank() == 0) {
    value = "Hello, World!";
  }
  broadcast(world, value, 0);
  std::cout << "Process #" << world.rank() <<
          " says " << value
          << std::endl;
  return 0;
}
```

**Figure 2 C++ broadcasting program as seen at [3]**

## 8 Distributed Architecture

In order to be useful, any distributed language should provide first-class features that reflect the program's runtime distributed environment. They can manifest themselves through particular fundamental data types and their operations, as well as through a runtime context. In order to make software development productive and programs runtime efficient, such architecture should reflect general functionality of the major components of modern networks. At the same time, it should be precise enough for the translator and runtime system to be able to map that architecture on the runtime environment, and to optimize it.

To be accepted among software engineers, the distributed primitives should augment, but not overwhelm the sequential language elements. These elements must remain its central part. Therefore, the Hello networking components are designed in order to blend with the main sequential core of the language, which resembles a simplified Java subset.

---

[2] Like the object mobility in Emerald [11], actors in Salsa [25], or process-orientation of Erlang [9].

The following list presents the essential elements of the Hello distributed architecture:

1) *Sourcepack* is a set of all source files with Hello programs under a single directory that together constitute a Hello package. Sources define hierarchies of related classes and interfaces. Each class represents both data that are instantiated as objects and functions which operate on objects at runtime. Interface defines shared data and function prototypes implemented by derived classes. Each source file must have extension `.hlo`, like `HelloWorld.hlo`.

2) *Runpack* is a shared library built from *all sources* of a source package by Hello translator: a sourcepack `abc` becomes runpack `abc.so`. Hello runtime engine loads and executes multiple runpacks at runtime. Engines from different computers exchange runpacks at runtime on demand, following the program control flow. Runpack code accesses data from memory partitions located on the same or different hosts. Runpacks contain machine instructions rather than intermediary code – they are binary libraries, which execute their code directly on CPU.

3) *Hello translator* is an executable `/usr/bin/het`, this translates Hello sourcepack into a runpack. The translator also archives and installs Hello sourcepacks.

4) *Computer* is a physical or virtual computing machine that may persistently store Hello packages in its directories, capable of running Hello packages, translator, hosts and runtime engines. A computer may reside on a network, which is used by runtime engines in order to transfer the Hello data, control flow, and runpacks between different hosts.

5) *Host* is a uniquely named set of runtime engines and partitions running on a computer. Each computer may run a single primary host and a number of secondary hosts. Engines from the same host directly access data from all partitions of their host. Engines from different hosts access each other's data through the network even if they run on the same computer.

6) *Partition* is a piece of virtual memory that holds shared runtime data – objects, as well as arrays of references to objects and primitive data. Each host may contain several partitions created by its engines. The partition data is available for multiple programs executing concurrently on either local or remote hosts.

7) *Runtime engine* is an executable `/usr/bin/hee`, this loads and executes runpacks. Runtime engine maintains its data heap accessible only to programs executed on its behalf. In addition, engines are responsible for automatic and transparent transferring of code, data and control flow between the hosts across the network. An engine accesses partitions from its host by mapping its virtual memory onto the partitions.

8) *Network* is a physical or virtual medium, which connects hosts from the same or different computers. Hello does not impose any particular operational characteristics on the network as long as the network is capable of maintaining locally unique computer addresses and of transferring code, data and control flow between the engines from the connected hosts. Hello is a protocol-agnostic language, as it imposes no requirements on the underlying network protocol; neither has it exposed any of the protocol elements to Hello programs[3].

9) *Queue* is a system object that queues up execution requests and subsequently dispatches them onto an execution thread; each Hello program executes on behalf of a thread designated to a particular queue.

10) *Event* is a call to a method with partially supplied arguments. Events accumulate on the queues, which execute events when subsequent calls complete the method's argument list. Events serve for delayed method invocation as well as for synchronization between local and remote queue threads.

11) **SID and Privilege** protect distributed data from unauthorized access. A Security Id (SID) is a 16-byte globally unique identifier that represents, via a bitmask, a set of privileges at runtime. Every queue has a set of pairs (SID, privilege) that determines operations allowed for the programs executed on behalf of the queue. In addition, an individual remote object may also have a set of such pairs to control operations allowed on the object. Finally, each host maintains a map of such pairs in order to determine the kinds of permissible remote requests executed on the host.

12) *Neighborhood* **and** *Path* assist in reaching distributed code and data across the network. A host neighborhood contains connected hosts that communicate directly, without any help from intermediary hosts. A host path is a sequence of connected hosts between two disconnected hosts; disconnected hosts communicate through the connected hosts from the path. The runtime engines build neighborhoods, accumulate host paths, and use them, transparently to Hello programs, while navigating the network between the hosts. Hello programs can control the neighborhood and paths formation as well as the path navigation policies.

# 9  Translation

The following figure illustrates how Hello translator `het` translates a source package into a runpack:

---

[3] However, the current runtime engine implementation requires the network to support the TCP protocol from IPV4.

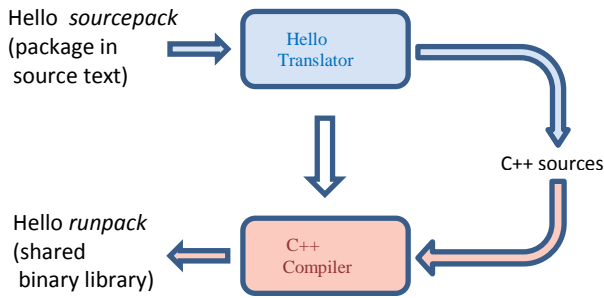

**Figure 3 Hello Translation**

Note that all source files from a given package participate in this operation. Although `het` can translate individual source files, it does not build the runpack unless it successfully translates all sources from the package. Another important feature of the Hello translation process is that `het` first translates Hello sources into C++ sources, and then automatically invokes the C++ compiler that further translates the C++ sources into a runpack – a shared binary library.

## 10 Sample Code – Remote Shell

The **Appendix** at the end of this paper presents a sample Hello code -- Remote Shell -- a program that executes a UNIX shell command on a remote host. Its code contains extensive comments in order to assist in the detailed understanding. In addition, here is a brief high-level explanation of its workings.

From the command line, the program accepts the name of the remote host, the number of buffers to parallelize data transfer, the UNIX command to execute on the remote host, and the arguments required for the execution. The program works in a remarkably simple way: it creates two instances of class `Shell` – one on the local host and one on the remote host. After that, the local engine passes command arguments to the remote instance, which executes the command. Then the local engine enters a loop that catches portions of the command's `stdout` output one after another and transfers it back to the local Shell instance. That instance dumps the data on the `stdout` of the local Hello runtime engine.

All three (potentially massive) data operations of reading data from the command `stdout`, transferring that data back to the original host, and dumping it on the `stdout` of the local host are performed in parallel. They involve Hello built-in asynchronous queuing operators and passing method parameters by value. This program also utilizes other distributed facilities such as creating instances of external classes on remote hosts, and synchronizing local expressions via queues. Together, these first-class distributed language features make Hello Shell simple to design and easy to develop while the resulting binary ends up being as efficient as its C counterpart `rsh`.

### 10.1    Hello, C!

It is fascinating to compare the amount of code from the Hello Shell program and the C code from the widely known UNIX remote shell rsh [24]. After deleting empty lines and all comments from the source code of both programs, running word count on both C `rsh.c` and Hello `Shell_World/Shell.hlo` produces this output:

```
hellouser@think:~/hem$ wc rsh.c
 390 1047 7606 rsh.c
hellouser@think:~/hem$ wc Shell.hlo
 116  427 2929 Shell.hlo
hellouser@think:~/hem$
```

Clearly, Hello Shell code is more than two times smaller than C rsh code. Obviously, some `rsh` code processes authentication and encryption while Shell code has no such functionality (although one can still tunnel Hello traffic through an SSL [22] proxy like `stunnel` [26]). However, even after deleting encryption and authentication from `rsh.c`, the word count still shows rsh code being significantly larger than Hello code:

```
hellouser@think:~/hem$ wc rsh.c.clear
 243  680 4724 rsh.c.clear
hellouser@think:~/hem$
```

## 11 Language Features

A full description of the Hello programming language, its translator, the runtime engine, and a number of working code examples are free and available for download from [14]. Here is the condensed digest of the language features.

The Hello distributed architecture had emerged from the practical experience of programming diverse distributed applications such as databases, cluster monitors, protocol discovery, email and video servers. Because of that, Hello design pursues the following two goals:

- The general-purpose distributed language should provide superior productivity for developing a variety of efficient and reliable distributed software.
- A mainstream programmer should be able to adopt this language in a short amount of time.

*Achieving these goals could facilitate the widespread transition from distributed tools, packages, and libraries to this distributed language.* Therefore, Hello combines the following unique properties, which set it apart from other distributed systems:

1) The first such property is the ability to create objects, access their data, and invoke their methods anywhere on the network, using simple Java-like syntax and semantics, which is also reminiscent of the RPC [23].
2) The second property is the language-embedded distributed architecture, whose elements directly represent the graph-structured underlying networking resources. These elements augment the single-host imperative language constructs, which represent the linearly structured components of a von Neumann machine.

3) In addition, Hello offers abstractions for distributed operations such as queued asynchronous and synchronous requests for remote synchronization, first-class bulk operations like group traversals and intelligent, deep copy algorithms, as well as asynchronous queued event processing.

Other Hello unique properties are safe data sharing in memory partitions and automatic transfer of packages across the network. Hello offers object-level security identifiers and privileges. It handles distributed failures via monitoring state label transitions, catching exceptions, and timeouts. In addition, Hello translator parallelizes distributed operations for runtime efficiency. Finally, Hello packages do not run inside a virtual machine. Hello translator converts them into shared binary libraries through the intermediary C++ invocation; the C++ translator can harness the power of its optimizer to optimize generated C++ code. Hello translator allows for embedding into Hello programs any C++ code and linking with any C++ library.

Hello programs are written inside object-oriented type hierarchies organized in packages. Hello translator translates Hello programs into C++ programs and then calls a C++ compiler to compile the generated C++ code into a 64-bit binary dynamic shared library. Hello runtime engine dynamically loads the libraries and executes them; it also transfers the libraries on demand across the network. Engines manage memory in partitions where the programs share runtime data. All this happens while executing concurrent threads under control of the queues. A runtime engine belongs to a host – a named collection of engines running on a single computer; any computer can run multiple hosts.

Hello programs manipulate strongly typed primitive data, arrays, strings, and class instances (objects). Local operations are performed on data from a thread stack, the engine heap and host partitions; remote operations are performed by navigating references to objects from other hosts. Method arguments and return values are transferred to any host by reference or value following an intelligent, deep copy algorithm. Remote and local objects assemble in groups for automated traversals. Hosts connect and engines transfer data, code, and control flow across the network transparently to the Hello programs. Hello programs can build neighbourhoods and paths of connected hosts and use them for better reliability, efficiency and security.

Currently, Hello v1.0.3 (alpha) is available on 64-bit Centos [6], Fedora [10] and Ubuntu [27] versions of the Linux OS [19] on Intel x86_64 architecture [17].

# 12 Conclusion

In conclusion, we discover a historical precedent for the emergence of the widely adopted distributed programming languages by drawing an analogy with the development of FORTRAN [12], which was the first high-level programming language that had alleviated the difficulties of binary programming. In addition, we rationalize the need for a widely accepted distributed language by drawing parallels between the difficulties of distributed and assembler software development [11]. Finally, we uncover the deeper currents carrying on its waves the distributed programming from the past -- through today -- into the future.

## 12.1   Historical Precedent

On the highest level, most general-purpose programming languages, in the context of a single computer, abstract the fundamental elements of software programs. Examples of these are code modules, data types, operations, and control flow. However, these languages do not provide first-class features for another significant programming component – automatic transfer of and access to code modules, data types, runtime data, operations, and control flow across the network between collaborating computers. Instead, they delegate such networking services to tools, libraries and packages outside the language proper.

In order to realize the importance of the first-class language features (i.e. those presented by language syntax and semantics), one should recall that programming languages serve as a bridge between a human understanding and the hardware realization of the algorithm. This bridge is extremely valuable: it provides a much easier path for human beings to control computers via high-level algorithmic abstractions rather than via low-level processor instructions. Ultimately, it saves precious time and money while developing software systems.

For example, the first high-level programming language FORTRAN developed in 1954--1957 at IBM [1], enjoyed tremendous success because it was able to express algorithms with the imperative control flow primitives, and centuries tested mathematical notation [15]. The language had freed programmers from the daunting burden of programming machine codes and auto-codes. The FORTRAN translator was able to translate program sources, check for programming errors, and compile efficient binaries from the program libraries without human intervention. At the same time, programmers, empowered with FORTRAN, were able to devote more time to solving application problems, instead of coding binary programs.

After FORTRAN invention, a plethora of programming languages had evolved [20]. However, despite immense richness of purposes and features, most of them share a single property inherited from the first language – they provide a bridge to only a single computer. Although many traditional languages allow for developing distributed applications, they still delegate network operations to extraneous tools, libraries and packages. This is simply because the languages lack the features to communicate over the network.

However, such extraneous software is complicated to use since its interface, being rich in complex low-level details, exposes the rigidity of the networking protocol to the application level. As a result, the quality of networking applications suffers because developers must spend significant efforts at a low-level coding in order to adhere to the subtleties of the networking interfaces. While the language itself has little or no knowledge of the networking operations, it is hard for a language translator to optimize network access; this, in turn, results in reduced efficiency and higher traffic of the networking applications.

As in 1950-s, today software engineers still lack an essential tool to address an exceptionally hard problem. While back then such problem was the low-level binary coding for a single computer, the today's problem is the low-level networking coding for distributed applications. Therefore, as the programming language FORTRAN solved the former problem more than half a century ago by providing first-class features to simplify single-host programming, another programming language might also solve the latter problem today. The solution could provide in the language the first-class features to simplify multi-host networking operations.

## 12.2  Technical Parallels

Anyone who had experienced binary or assembler programming and, in addition, had been involved in a distributed software project, can think of the binary coding problems facing programmers of early 1950's being quite similar to the networking coding problems of today. The following list enumerates the exact parallels:

1) <u>*A well-defined central processing element*</u>: a CPU versus a server on the network;
2) <u>*A program that needs service from the central element*</u>: an assembler, auto-code or binary program in memory, executing CPU instructions versus a client program on the network, submitting server requests;
3) <u>*A hard to use interface*</u>: assembler language, auto-code or binary code versus networking interface;
4) <u>*A pattern of failure*</u>: crash of CPU or server usually causes termination of assembler or client program;
5) <u>*Programming, debugging, and optimization*</u>: in both cases, techniques are manual and complicated [11], [28].

Obviously, it should be of no surprise that a similar solution could solve a similar problem. In other words, a distributed programming language might improve networking programming today the same way FORTRAN had improved binary coding more than half a century ago.

## 12.3  Connecting Past and Future

During millennia-long mathematical development [16] and centuries-long technological progress [7], human beings performed calculations individually. When needed, they used to carry out computational and mathematical interactions through inter-personal collaboration, education, written correspondence, manuscripts, books and magazines. It is only recently, after the invention of the computer networks some 50 years ago [8], the automated distributed computations had become possible. Therefore, it is not surprising that the area of distributed programming is less mature than the field of single-host programming, which takes its roots in the individual computational techniques evolved through thousands of years.

One might speculate that it might take more centuries for human civilization to develop (yet unknown) automated paradigm for distributed computations. However, it appears that the movement towards that goal can be started today. Its first step should abandon a hodgepodge of single-host languages and distributed tools in favour of widely accepted general-purpose distributed programming languages.

## Appendix – Remote Shell Source

```
10  // each source must begin with
11  // the package directive
12  package Shell_world;
13
14  // this class is declared external because
15  // its instances can be created on a remote host
16  // and also because once created they can
17  // be accessed from a remote host
18  external class Shell
19  {
20      // this is constructor of Shell class
21      external public Shell() {}
22
23      // entry point to the program -- must
24      // be called main; accepts command line
25      // arguments as an array of strings
26      static public int main(char [][]argv)
27      {
28          // built-in method sizear() returns
29          // the number of elements in an array
30          int argc = sizear(argv, 1);
31
32          // quit program if not enough arguments
33          if ( argc <= 1 )
34              return 0;
35
36          // first argument must be a host
37          // name were to execute command
38          // specified on the Shell command line;
39          // the built-in method hello() returns a
40          // reference (ref for short) to a host on
41          // the network with the given name
42          host hst = hello(argv[0]);
43
44          // if host not found then quit
45          if ( hst == null ) {
46              #C { cout << "host not found\n"; }
47              return -1;
48          }
49
50          // the 'create' expression is similar
51          // to 'new' expression, except it creates
52          // an instance not in the engine heap
53          // but in a partition from the specified
54          // host; if the host is remote, then an
55          // instance of class Shell is created
56          // on that host
57          Shell shl = create (hst) Shell();
58
59          // this line executes a method run() from
60          // class Shell on the just created
61          // remote object referred to by ref shl;
62          // after run() completes, this program exits
63          shl.run(argv, this_host);
```

```
 64      }
 65
 66      // this defines the size of a buffer for data
 67      // to be transferred from the stdout of the
 68      // command executed on the remote host to
 69      // the local host -- this data will be printed
 70      // on stdout of the program main() running on
 71      // the local host
 72      enum { BUFSIZE = 1024 * 1024 * 4 }
 73
 74      // this method executes on the remote host:
 75      // it accepts command and its parameters
 76      // in the character array [][]argv, and a ref to
 77      // the remote host 'back' -- the host from where
 78      // remote Shell command has been started
 79      public external void run(copy char [][]argv,
 80                                       host back) {
 81
 82          // get the count of arguments
 83          int argc = sizear(argv, 1);
 84
 85          // -the first argument is the name
 86          //  of this remote host;
 87          // -the second argument
 88          //  specifies how many buffers to fill in
 89          //  with the stdout data in parallel:
 90          //  this argument is for experimentation,
 91          //  just to see how the speed of data
 92          //  transfer depends on the amount of
 93          //  buffers being filled in parallel
 94          // -the third argument is the remote
 95          //  command name
 96          // -the rest are the arguments for
 97          //  the remote command
 98          //
 99          // If no command is specified then quit
100          if ( argc <= 2 || back == null )
101              return;
102
103          // set the count of buffers from
104          // command line
105          int BUFCNT;
106          char []bc = argv[1];
107          #C { $BUFCNT = atoi($bc().__adc()); }
108          if ( BUFCNT < 0 )
109              BUFCNT = 0;
110
111          // create another Shell instance, this time
112          // on the source host from where the Shell
113          // has been launched
114          Shell rso   = create (back) Shell();
115
116          // on that same host create a queue
117          // for dumping command stdout output on
118          // the Shell command output in parallel
119          // with accepting the next portion of
120          // stdout from the remote host
121          queue rsq   = create (back) queue();
122
123          // create a queue on this host for
124          // sending portions of stdout back to the
125          // source host in parallel with accepting
126          // stdout from the command stdout
127          queue rtq   = create queue();
128
129          // create a character array and
130          // fill it with the command and its
131          // arguments using operations
132          // concatenation + and +=
133          char []line = create char[0];
134          for ( int i = 2; i < argc; i++ )
135              line += argv[i] + " ";
136
137          // also append redirection of stderr
138          // to make sure that any error messages
139          // are transferred back to the source host
140          line += "2>&1";
141
142          // create buffers in shared memory:
143          // the count of buffers 'buc' is set from
144          // command line, the size of each buffer
145          // is set from the enum BUFSIZE
146          int buc = BUFCNT?BUFCNT:1;
147          char [][]output = create char[buc][BUFSIZE];
148
149          // this is command exit code
150          int res    = 0;
151
152          // this holds the length of data
153          // in the buffer
154          int len    = 0;
155
156          // this indicates if all data has been
157          // received from the command stdout
158          bool done = false;
159
160          // this executes the command using C++ code:
161          // enclosed in the embedded block denoted
162          // by '#D' -- a UNIX system call popen()
163          // is called and the command with all
164          // arguments is launched; a file descriptor
165          // 'pipe' is filled by this call: it is
166          // used below to receive the command stdout
167          #D {
168              FILE *pipe = popen($line().__adc(),
169                                         "r");
170              if ( pipe == NULL ) $done = true;
171          }
172
173          // now the program enters a loop that reads
174          // command stdout; counter 'i' indicates
175          // which buffer to use; boolean 'done'
176          // is set inside the loop when all
177          // stdout data has been received
178          int i = 0;
179          while ( !done )
180          {
181              // check if all buffers have been used
182              if ( i >= BUFCNT )
183              {
184                  // if all buffers are filled in
185                  // by stdout data, then the
186                  // constant expression '1' is
187                  // evaluated on the local queue
188                  // 'rtq' using the queued
189                  // expression evaluation
190                  // operator '<=>'; this operator
191                  // places an expression at the
192                  // end of the queue -- when
193                  // that expression reaches the
194                  // head of the queue then control
195                  // returns back to the current
196                  // thread
197                  rtq <=> 1;
198
199                  // because queue 'rtq' contained
200                  // requests to send data to the
201                  // remote source host, control
202                  // reaches this point only after
203                  // all such requests have
204                  // finished; then the current
205                  // buffer number is reset back
206                  // to 0
207                  i = 0;
208              }
209
210              // get next buffer and advance
211              // buffer count
212              char []outbuf = output[i++];
213
214              // remember the address of that buffer
215              #D { char *b = $outbuf().__adc(); }
216
217              // this is the count of bytes
218              // left to read into the buffer
219              // from command stdout
220              int toread = BUFSIZE;
221
222              // keep reading from command stdout
223              // until this buffer is full
224              #D {
225                  char *o = b;
226                  while ( $toread > 0 ) {
227                      $len = fread(o, 1, $toread,
228                                          pipe);
229                      $toread -= $len;
230                      o += $len;
231                      if ( feof(pipe) ) {
232                          $done = true, $res = 0;
233                          break;
234                      }
235                      else if ( ferror(pipe) ) {
236                          $done = true, $res = -1;
237                          break;
238                      }
239                  }
240              }
241
242              // this is the count of bytes read
243              // from the command stdout; it is
244              // supposed to be equal to BUFSIZE
245              // for all buffers except possibly
246              // the last one for which it can
247              // be less than BUFSIZE
248              len = BUFSIZE - toread;
249
250              // if no buffers were ordered then
251              // do not transfer any data back
252              // to the source host; otherwise
253              // send the method message named
254              // 'send()' to 'this' Shell object
255              // on the queue 'rtq'; pass the data
256              // buffer and the length, and the
257              // refs to the remote Shell object
258              // 'rso' and to remote queue 'rsq',
259              // and the completion indicator 'done'
260              if ( BUFCNT == 0 )
261                  ;
```

```
262                else
263                    rtq #> (this, send(rso, rsq, outbuf,
264                                       len, res, done));
265                // at this point the sent message 'send'
266                // begins the transfer of the
267                // accumulated data from 'outbuf'
268                // back to the source host via a
269                // invocation of external method
270                // 'rcv' which has a 'copy' parameter
271                // 'data' to which 'outbuf' buffer
272                // is passed as an argument; meanwhile,
273                // this loop continues to receive the
274                // rest of command stdout
275            }
276            // this queued constant expression
277            // evaluation is needed to wait until
278            // remote source Shell object completes
279            // its work, so that this target Shell
280            // object exits afterwards; premature exit
281            // of the target Shell will result in an
282            // exception on the source shell
283            rtq <=> 1;
284        }
285
286        // this method 'send' is declared as
287        // 'message' so that it can be placed on
288        // the queue for asynchronous execution
289        public message void send(Shell rso, queue rsq,
290                                 char []data, int len,
291                                 int res, bool done)
292        {
293            // after this message reaches the head
294            // of the queue it has been placed upon,
295            // it calls method 'rcv' on the remote
296            // source Shell instance 'rso'; the buffer
297            // 'data' is copied remotely by the
298            // runtime engine back onto the
299            // remote source host
300            rso.rcv(rsq, data, len, res, done);
301        }
302
303        // this method is declared 'external' because
304        // it is called remotely from the target host
305        // on a shell object from the source host;
306        // parameter 'data' is declared 'copy'
307        // for passing array of data by value across
308        // the host
309        public external void rcv(queue rsq,
310                                 copy char []data,
311                                 int len,
312                                 int res,
313                                 bool done)
314        {
315            // when the source host receives the
316            // command stdout in the 'copy' parameter
317            // 'data', it sends method 'dump' to
318            // 'this' object (which is actually 'rso'
319            // on the source host; method 'dump'
320            // actually writes command stdout from
321            // remote target host on the local
322            // source host)
323            rsq #> (this, dump(data, len, res, done));
324
325            // when all data has been received, this
326            // queued expression waits until
327            // all output completes to avoid
328            // abrupt interrupt of the output
329            if ( done )
330                rsq <=> 1;
331        }
332
333        // this method 'dump' just prints
334        // the data received from remote target host
335        // on stdout of the runtime engine from
336        // the local source host
337        public message void dump(char []data,
338                                 int len,
339                                 int res,
340                                 bool done) {
341            // output data using embedded C++ block
342            #C {
343                char *buf = $data().__adc();
344                int wrt = 0;
345                while ( $len > 0 ) {
346                    wrt = write(1, buf, $len);
347                    if ( wrt == -1 )
348                        break;
349                    len -= wrt, buf += wrt;
350                }
351            }
352
353            // if exit code indicated a failure (!=0)
354            // then dump that exit code
355            if ( done && res )
356                #C { cout << "exit code " << $res; }
357        }
358 };
```